\documentclass[twocolumn,
pra,
 showpacs,floats,preprintnumbers,amsmath,amssymb]{revtex4}
\usepackage{epsfig}
\usepackage{amsmath}
\renewcommand{\(}{\left(}
\renewcommand{\)}{\right )}
\renewcommand{\[}{\left [}
\renewcommand{\]}{\right ]}
\def\fslash#1{#1 \!\!\! \slash}
\def\pa{\partial}

\def\varp{\varepsilon}
\def\bea{\arraycolsep .1em \begin{eqnarray}}
\def\eea{\end{eqnarray}}

\def\vp{{\bf p}}

\def\vk{{\bf k}}

\def\Tr{{\rm Tr}}

\let\be=\beta
\let\no=\nonumber

\def\eq#1{Eq.(\ref{#1})}
\def\refr#1{\cite{#1}}

\def\eqs#1{Eqs.(\ref{#1})}
\def\s0#1#2{\mbox{\small{$ \frac{#1}{#2} $}}}
\def\0#1#2{\frac{#1}{#2}}

\def\anp#1#2#3{Adv.\ Nucl.\ Phys. \ {\bf #1}, #2 (#3)}
\def\plb#1#2#3{Phys. Lett. {\bf B #1}, #2 (#3)}

\def\pra#1#2#3{Phys. Rev.  {\bf A #1}, #2 (#3)}
\def\prb#1#2#3{Phys. Rev.  {\bf B #1}, #2 (#3)}
\def\prc#1#2#3{Phys. Rev.  {\bf C #1}, #2 (#3)}

\def\prl#1#2#3{Phys. Rev. Lett. {\bf #1}, #2 (#3)}
\def\ann#1#2#3{Ann. Phys. (N.Y.) {\bf #1}, #2 (#3)}
\def\anp#1#2#3{Adv. Nucl. Phys. {\bf #1}, #2 (#3)}

\def\epl#1#2#3{Europhys.\ Lett.{\bf #1}, #2 (#3)}
\def\prep#1#2#3{Phys.\ Rep.\ {\bf #1}, #2 (#3)}

\def\ijmpe#1#2#3{Int.\ J.\ Mod.\ Phys.\ {\bf E #1}, #2 (#3)}

\def\ibid#1#2#3{{\it ibid.}, {\bf #1}, #2 (#3)}
\def\sci#1#2#3{Science\ {\bf #1}, #2 (#3)}
\def\nat#1#2#3{Nature\ {\bf #1}, #2 (#3)}
\begin{document}
\title{$D$-dimensions Dirac
fermions BEC-BCS crossover thermodynamics}
\author{Ji-sheng Chen\footnote{chenjs@iopp.ccnu.edu.cn}}
\address{%$^a$
Institute of Particle Physics $\&$ Physics Department, Hua-Zhong
Normal University, Wuhan 430079, People's Republic of China}
\begin{abstract}
An effective Proca Lagrangian
    action is used to address the vector condensation Lorentz violation effects
    on the equation of state of the strongly interacting fermions system.
The interior quantum fluctuation effects are incorporated  as an
    external field approximation indirectly through a fictive generalized Thomson
    Problem counterterm background.
The general analytical formulas for the $d$-dimensions
    thermodynamics are given near the unitary limit region.
In the non-relativistic limit for $d=3$, the universal
    dimensionless coefficient
    $\xi ={4}/{9}$ and energy gap $\Delta/\varepsilon_f  ={5}/{18}$ are
    reasonably consistent with the existed theoretical and
    experimental results.
In the unitary limit for $d=2$ and $T=0$, the universal
    coefficient can even approach the extreme occasion
    $\xi=0$ corresponding to the infinite effective fermion mass $m^*=\infty$ which can be mapped to the
    strongly coupled two-dimensions electrons and is quite similar to
    the three-dimensions Bose-Einstein Condensation of ideal boson gas.
Instead, for $d=1$, the universal coefficient $\xi$ is negative,
    implying the non-existence of phase transition from superfluidity
    to normal state.
The solutions manifest the quantum Ising universal class
characteristic of the
    strongly coupled unitary fermions gas.
\end{abstract}


\maketitle
\section{Introduction}

 Since DeMarco and Jin achieved the Fermi
degeneracy\refr{DeMarco},
    the property of the strongly coupling fermions system under the ultra-cold extreme
    conditions serves as a bewitching topic in the fundamental
    Fermi-Dirac statistical physics.

The universal thermodynamics properties learned from the
    ultra-cold atomic experiments can be related to many other
    realistic many-body physics problems\refr{Bertsch1999,Baker}.
For example, in nuclear physics, the magnitude of the
    neutron-neutron  $S$-wave scattering length $a_{NN}=-18.6$ fm is
    much larger than the interaction range given by $R\approx 1/m_\pi \sim 1.4 $ fm,
    which will indicate the low energy thermodynamics universal
    property in the dilute regime\refr{Heiselberg2000}.
Furthermore, the Feshbach resonance physics can be also
     related with the strongly interacting $SU(N_c)$ \textsl{non-Abelian} plasma low
    viscosity physics realized in the ultra-relativistic heavy-ion
    collisions or existed in the core of neutron
    star\refr{Schaefer} and even the ``high-temperature" $T_c$
    physics\refr{chenq2005}.

Across the Feshbach resonance regime, the interaction changes from
weakly to strongly
    attractive  according
    to the magnitude of the applied external magnetic field  $B$.
At this Bose-Einstein
    Condensation (BEC)/Bardeen-Cooper-Schrieffer (BCS) crossover point,
    the scattering length diverges $|a|=\infty$ due to the existence of a zero-energy bound state.
This characteristic leads to the universal thermodynamics,
    i.e., the system details do not contribute to the thermodynamics
    properties.
The intriguing universal unitary fermions thermodynamics
properties attract  much attention recently in the
literature\refr{Heiselberg2000,Schaefer,chenq2005,physics/0303094,
Astrakharchik2004,cond-mat/0404687,Bulgac2005,Ho2004}.

In the dilute unitary limit, the important physical quantities
energy density $E/N
    = \xi\(E/N\)_{ideal} $ and energy gap $\Delta$ or critical phase
    transition temperature $T_c$,
    are the two key parameters to be determined experimentally and calculated theoretically.
The dimensionless proportionality constant $\xi$ is called the
    universal factor.
The $\Delta $ is also proportional to the Fermi kinetic energy.
    Many growing theoretical and experimental efforts about the
    universal factor $\xi$ have been made in recent years
    Refs.\refr{Heiselberg2000,physics/0303094,Astrakharchik2004,cond-mat/0404687,Bulgac2005,Ho2004}\refr{Gehm2003}.
It is worthy noting that the differences for the critical
    temperature or the energy gap can be as large as several times
    among the theoretical or experimental results existed in the
    literature.
Up to now, there is not a reliable analytical method to deal with
the unitary limit problem although there are many updating
theoretical analytical attempts. Beyond the naive perturbative
expansion techniques or the lowest
    order mean field theory, a novel method is to be explored for
    the unitary limit fermions thermodynamics\refr{Pethick}.
The essential task is how to incorporate the in-medium nonlinear
    quantum fluctuation/correlation effects into the thermodynamics in
    a reasonably controllable way.

Recently, inspired by the homology between this unitary limit
    topic and the nonperturbative long range infrared singularity
    manifested by the hot and dense gauge field theory, we make an
    analytical detour to attack this intriguing
    problem\refr{chen2006}.
Our initial observation is that the scattering cross section
between the two-body particles at the zero energy bound unitarity
is limited as $S$-wave $\sigma = 4\pi /k^2$(with $k$ being the
relative wavevector of the colliding particles), while the
    gauge propagator is $\Delta _{\mu\nu}\sim g_{\mu\nu}/{k^2}$.
The interesting analytical results $\xi =4/9$ and
    $\Delta/\varepsilon_f =5/18$
    comparable with other theoretical ones are obtained by taking
    the still unsolved classical Thomson Problem as a potential
    nonperturbative quantum many-body arm.
In the non-relativistic limit with $T=0$, the analytical
expression for the energy density is the same as
    that obtained by Steele with the power counting in the effective
    theory framework\refr{Steele}.

The relativistic continuum quantum field theory formalism
provides a natural framework to address the infrared long range
singular correlations associated with the density of state in
many-body physics according to quantum statistical physics. In
terms of the relativistic formalism we can certainly achieve more
than those we
 need in a non-relativistic theory, but we can recover
all the non-relativistic limit physical quantities by expanding in
powers of $k_f/m$ at any time especially for the $T=0$ universal
unitary gas thermodynamics. In this work, we will continue the
relativistic continuum field theory attempts
    for the strongly interacting
    Fermi-Dirac statistical physics in terms of the in-medium Lorentz
    violation.
After giving the general analytical expressions for the energy
    density functional and pressure as well as entropy density
    according to the density functional theory near the unitary limit,
    the final analytical results comparable with other theoretical
    approaches will be obtained by applying the corresponding Taylor
    series expansion.

Initialed with the recent discussions about the low $d$-dimensions
unitary fermions gas\refr{Nussinov},
    we extend the Thomson Problem  counterterm analytical method to arbitrary $d$-dimensions with the relativistic continuum field theory formalism.
The $d=2$-dimensions solution reflects the essential interesting
physics. The universal dimensionless coefficient $\xi $ can
    approach to the specific value $\xi =0$,
    corresponding to the two-dimensions strongly coupled electrons with the infinity effective mass $m^*=\infty $.
This is also very similar to the well-known BEC phenomena for the
    three-dimensions ideal boson gas with the simultaneous vanishing of
    both the energy density, pressure and entropy density at $T=0$.
Furthermore, the energy gap can be zero. Instead, for the
    one-dimension,
    the $\xi$ can be negative, corresponding to non-existence of the
    phase transition from the superfluidity to normal state at $T=0$.
The properties of these solutions manifest the strongly coupled
    quantum Ising universal class characteristic of the unitary fermions
    thermodynamics realized in the artificial experimental environments.

The present work is organized as follows. In section \ref{sec2},
    the adopted low energy effective quantum field theory formalism of the relativistic
    continuum
    Proca Lagrangian and corresponding Thomson Problem counterterm nonperturbative approach for the
    strongly interacting unitary gas thermodynamics are described.
The effective potential and the general thermodynamics quantities
    for finite scattering length $a$ as well as the concrete
    comparisons with existed results are presented in section \ref{sec3}.
Moreover, the $d$-dimensions universal thermodynamics at unitarity
    are discussed in section \ref{sec4} and the concluding remarks are
    made in the final section \ref{sec5}.

Throughout this work, the natural units with $c=\hbar =k_B=1$ are
adopted.

\section{Formalism}\label{sec2}
\subsection{Effective interaction action}

A nonperturbative approach is
    crucial to understand the strongly correlating dense and hot
    many-body physics.
Near the strongly interacting unitary limit, any form of effective
    interaction can be used to detect the universal thermodynamics
    properties in the pseudo-potential spirit of statistical
    physics.
We extend previous work\refr{chen2006} to near unitary limit
regime motivated by the successful relativistic quantum
    many-body formalism\refr{walecka1974}.
To model the strongly interacting fermions thermodynamics in terms
    of the fundamental in-medium Lorentz violation effects, let us
    consider the relativistic Dirac fermions interacting with an
    auxiliary Proca-like Lorentz massive vector field\refr{Proca}
\bea\label{original}
    {\cal L}_{\textsl{effective}}=&&{\bar \psi} (i \gamma _\mu\pa ^\mu -m) \psi\no\\
    && -\014 F_{\mu\nu} F^{\mu\nu}+\012 M_A^2 A_\mu A^\mu+A_\mu
    J^\mu \no\\
    &&+\delta {\cal L}_{\textsl{Auxiliary~~Thomson~Counterterm~Background}}.
\end{eqnarray} The $A_\mu$ is the vector field with the field stress
\bea
    F_{\mu\nu}=\pa_\mu A_\nu-\pa_\nu A_\mu.
\eea

In the effective action ${\cal L}_{I, A-\psi}=\012 M_A^2 A_\mu
A^\mu+A_\mu
    J^\mu$, the vector
current
    contributed by the fermions is
\bea
    J^\mu = g {\bar \psi} \gamma ^\mu \psi.
\eea

In \eq{original}, $m$ is the bare fermion mass with the vector
    boson mass squared assembling both a bare $m_A^2$ and a
    fluctuating one $ m_{\tilde{B}}^2$\refr{Alfaro2006}, $M_A^2=m_A^2 +
    m_{\tilde{B}}^2$.
Here, a bare Lorentz violation term $1/2 m_A^2 A_\mu A^\mu$ is
    included obviously,
    which determines the interaction strength in vacuum with the bare electric charge
    $g$.
For example, the $S$-wave scattering length $a$ is given through
    the low energy Born approximation in the language of the low energy quantum
    mechanics in three-dimensions
\bea\label{scattering} \0{g^2}{m_A^2}=\0{2\pi a}{m}.\eea

\subsection{Induced interaction determined by Thomson counterterm background}
The induced Lorentz violation term caused by the Thomson
background $1/2 m_{\tilde{B}}^2 A_\mu A^\mu$ represents the
many-body
    renormalization effect, which has been introduced with a
    hidden local symmetry manner(\textit{Abelian} Higgs model) by mapping
the unitary limit topic to the infrared singularity correlation in
dense and hot gauge theory\refr{chen2006}. Taking the coupling
constant $g$ to be an ``electric" charge of
    the fermions,
    makes it possible for us to introduce
    the generalized renormalization condition,
    i.e., the fictive Thomson stability condition.
With the physical constraint Thomson stability condition,
    the parameter $ m_{\tilde{B}}^2$ is found to be the negative of
    the gauge invariant Debye mass squared in the unitary limit
    $|a|=\infty$\refr{chen2005,chen2006}.
The gauge invariant Debye mass associated the density of state
(DOS) reflects the essential quantum fluctuation effect on the
thermodynamics.

More generally, the analytical expression for the Debye mass
squared is the gauge invariant static infrared limit of the $00$
ingredient for the vector boson polarization tensor $ \Pi
^{\mu\nu}_{A}(p_0,\vp)$, which can be also calculated with the
full fermion
    propagator through the Dyson-Schwinger equations or relativistic
    random phase approximation (RPA)\refr{chen2002}
\bea\label{self-energy}
    \Pi ^{\mu\nu}_{A}(p_0,\vp)=g^2T\sum _{k_0}\int_k \Tr
        \[\gamma ^\mu \0{1}{\fslash{k}-m}\gamma
        ^\nu\0{1}{(\fslash{k}-\fslash{p})-m}\],\\
               m_{\tilde{B}}^2=\Pi^{00}_A(0,|\vp|\to 0)=-\Pi^{L}_A(0,|\vp|\to 0).\no \eea
In \eq{self-energy}, the $0$-component of the $d+1$-dimensions
    momentum $k=(k_0,\vk )$ in the fermion loop is related to temperature T and
    \textsl{effective} chemical potential
    $\mu ^*$ via $k_0=(2 n+1) \pi T i +\mu ^*$ determined by the pole of the full fermion propagator
    with the relativistic Hartree approximation (RHA) formalism.

    Throughout this paper,
    the shorthand notation $\int _k=\int {d^d{\bf k}}/{(2\pi
    )^d}$ will be adopted for the $d$-dimensions
    abstract Dirac phenomenology\refr{peskin1995,Tsvelik}.
With finite temperature field theory, the general analytical
    expression for $d$-dimensions unitary fermions gas is derived
    to be
\bea\label{debye-mass-relation}
   \0{m_{\tilde{B}}^2}{g^2}&&=- 2\int _k
  \0{(d-2) E_k^2 +\vk^2 }{ E_k \vk ^2}\[f +\overline{f}\].
\eea In the above expressions, \bea
     f=\01{e^{\be (E_k-\mu ^*)}+1},~~~~
    \overline{f}=\01{e^{\be (E_k+\mu ^*)}+1},
\eea are the distribution functions for (anti-)particles with $E_k
    =\sqrt{{\bf k}^2+m^2}$ and  $\be =1/T$ the inverse temperature.

\begin{figure}[ht]
        \centering
        \psfig{file=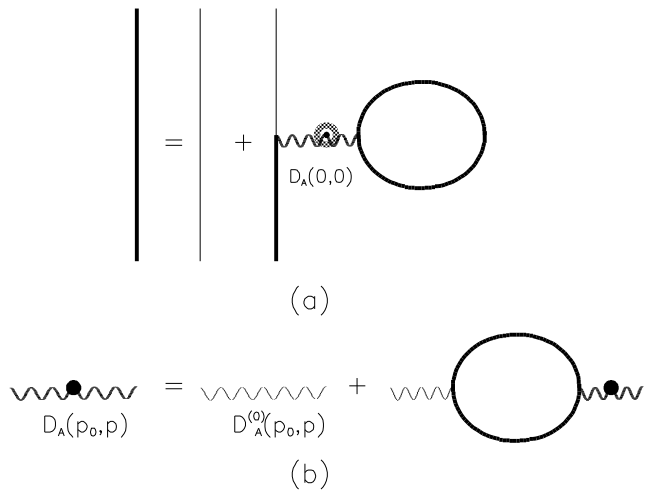,width=8.5cm,angle=-0}
        \caption{
        \small
The generalized coupled Dyson-Schwinger equations incorporating
the Landau pole contributions:
 (a), The full fermion
propagator determined with the effective
    interaction through the in-medium \textsl{anti-screened} vector boson
    ``propagator" in instantaneous approximation
    $\propto 1/(m_A^2-\Pi^L_A(0,|\vp|\to 0))$;
(b), The full vector boson propagator calculated with the
in-medium full fermion propagator through RPA. }\label{fig1}
\end{figure}

%
%In three-dimensions, the relation \eq{debye-mass-relation} is
%reduced to \bea\label{debye-mass-relation}
%    m_{\tilde{B}}^2&&=- \0{g^2}{\pi^2} \int _0^\infty d|\vk|
%    \0{(2\vk ^2+m^2)}{E_k}\[f +\overline{f}\].
%\eea
This kind of calculations can be indicated by the comprehensive
    diagrammatic representation as indicated in Fig.\ref{fig1}.

\subsection{Diagrammatic expression for the coupled generalized Dyson-Schwinger
equations}

It should be stressed the in-medium boson propagator with the
\textsl{negative} of the gauge invariant longitudinal component
    $\Pi_A^L(0,|\vp|\to 0)=-\Pi_A^{00}(0,|\vp|\to 0)$ of the polarization tensor
    $\Pi_A^{\mu\nu}(0,|\vp|\to 0)$ presented in Fig. 1.a.
This minus sign implies that the induced interior correlation
    contribution is realized with the external field approximation formalism
    indirectly caused by the opposite charged Thomson counterterm background.

The \textsl{minus} sign in front of $\Pi^L_A(0,|\vp|\to 0)$ is
    crucial,
    which ensures the theoretical thermodynamics self-consistency and
    the unitarity of the final physical results in the unitary
    limit\refr{chen2005}.
The comprehensive Landau dynamical screening effects are
    automatically incorporated while the concrete physical contents
    are quite different from the conventional weak coupling resummation
    techniques.
With these two Feynman diagrams, the relevant algebra equations
formalism instead of the coupled integral equations one makes it
possible for us to deal
    with the low energy density and temperature
    fluctuation/correlation effects on
     the strongly interacting fermions thermodynamics in an
    accurate way.
In other words, the in-medium density correlation effects are
    included through the effective chemical potential instead of
    directly through the effective fermion mass.
The spirit hidden in this kind of approach is quite similar to the
    Kohn-Sham density functional theory\refr{Kohn1965,walecka1974}.
Especially, the integral momentum divergence is naturally gauged
    by the relativistic kinematic factors due to taking the density
    functional theory formalism to approach the universal unitary gas thermodynamics.

That the correlation effects are embodied through the fictive
    opposite electric charged Thomson counterterm background can also avoid the
    theoretical double counting trouble, which leads to the key \textsl{minus} sign difference as
    found in previous works\refr{chen2005}.
    This minus sign difference is inspired by the anomalous thermodynamics discussions of neutron-star nuclear
matter with either short or long-range force
interactions\refr{gulminelli2003,note}.

\subsection{Parameters $m_{\tilde{B}}^2$ at $T=0$}
At $T=0$, from the general expression \eq{debye-mass-relation},
one can have\bea \0{m_{\tilde{B}}^2}{g^2}=-\0{2
d}{(2\sqrt{\pi})^{d}}\0{k_f^{d-2}E_f}{\Gamma (\0d2 +1)}, \eea
    which can be reduced to
$\0{m_{\tilde{B}}^2}{g^2}=-\0{2
d}{(2\sqrt{\pi})^{d}}\0{k_f^{d-2}m}{\Gamma (\0d2 +1)}$ for the
non-relativistic limit. The fermion density in $d$-dimensions is
\bea n =\02{(2 \sqrt{\pi
    })^d} \0{k_f^d}{\Gamma(\0d2 +1)}.
\eea In above expressions, the $\Gamma
    (\0d2 +1)$ is the Gamma function.

 For example, with $d=3$, the parameter $m_{\tilde{B}}^2$ is \bea
    \0{m_{\tilde{B}}^2}{g^2}%=-\0{g^2}{\pi ^2}k_f
    %\sqrt{m^2+k_f^2}
    =-\0{k_fE_f}{\pi ^2},
\eea with $k_f$ being the Fermi momentum and
    $E_f=\sqrt{k_f^2+m^2}$ the relativistic Fermi energy.
In the
    non-relativistic limit, the corresponding result is $\0{m_{\tilde{B}}^2}{g^2}=-\0{k_fm}{\pi ^2}$.

For $d=2$, one can have $\0{m_{\tilde{B}}^2}{g^2}=-\0{E_f}{\pi }$
with the expression \eq{debye-mass-relation}. The corresponding
non-relativistic result is
    $\0{m_{\tilde{B}}^2}{g^2}=-\0{m}{\pi }$.
For $d=1$, the result is $\0{m_{\tilde{B}}^2}{g^2}=-\0{2E_f}{\pi
k_f}$, while the non-relativistic limit is
$\0{m_{\tilde{B}}^2}{g^2}=-\0{2m}{\pi
    k_f}$.
In the unitary limit, they are negative of the gauge invariant
Debye/Thomas-Fermi mass squared\refr{Tsvelik}.

Before making the following calculations, let us discuss the
    physical meanings of the coupling constants represented by the vector boson mass $m_A$ and electric
    charge $g$.
They are the parameters introduced to characterize the bare vacuum
    interaction strength (scattering length $a$ in three-dimensions) between particles
    with relativistic continuum Dirac formalism.
Only in the three-dimensions, the coupling constant $g$ can be of
dimensionless, which can be clearly seen
    from \eq{scattering}. Unusually, it will have the mass
    dimensional $\mu ^{\0{3-d}2}g'$, where $g'$ is dimensionless and
    $\mu$ is an arbitrary mass scale of the problem.
%In other words, the interaction strength is characterized by the
%    set of $m_A$ and $g$.
%, i.e., corresponding to the dimensionless
%    coupling constants  $C_\om ^2=m^2 g_\om^2/m_\om^2$ etc. of the relativistic nuclear
%    theory in three-dimensions\refr{walecka1974}.

\subsection{Effective interaction  strength}

Physically, the interaction between the fermions is renormalized
    by the Lorentz violating many-body
    environment and the induced interaction strength between the particles is
    characterized by the three parameters $m_A^2$, $m_{\tilde{B}}^2$
    and $g^2$. The in-medium effective
     scattering length formalism can give relevant concise expressions to describe
     the strongly correlating effect in a nonperturbative manner\bea
    a_{eff}=\0{m}{2\pi }\0{g^2}{M_A^2}=a\01{1+\0{m^2_{\tilde{B}}}{m_A^2}}.\eea

It is worthy noting that the $m^2_{\tilde{B}}$ and $g^2$ appear in
the analytical expressions as a fraction ratio. Actually, the
magnitude of this fraction ratio $m^2_{\tilde{B}}/g^2 $
characterizes the remarkable low energy long range infrared
quantum fluctuating effects associated with the DOS on the
    universal thermodynamics  according to
    the quantum statistical physics. Therefore, there are not any
    remained adjustable or expansion parameters, i.e., the physical
    vacuum interaction strength parameter/$S$-wave scattering length $a$ controls
    completely the final thermodynamics quantities as given below.

\section{Effective potential through Lorentz vector condensation with external field
approximation}\label{sec3}

In methodology, the quantum many-body correlation/DOS
    effects are incorporated as an external field approximation to
    ensure the theoretical thermodynamics self-consistency,
    from which the effective potential can be obtained with the standard path integral
    by taking the relativistic Hartree instantaneous approximation.
The external field approximation formalism implies that the
    vector current conservation/gauge invariance can be guaranteed by the Lorentz
    transversality condition realized by RHA
\bea
    \pa _\mu A^\mu=0,
\eea from which the effective potential
reads\refr{walecka1974,kapusta1989} \bea
   \label{potential}
   \Omega/V&&=-T\ln Z\no\\
       &&=-\012 M^2_A A_0 ^2 -T  \int _k \sum _i
        \[\ln (1+e^{-\beta (E_{k,i}-\mu_i^*)})\right.\no\\
       &&
       ~~~~~~~~ \left.+\ln (1+e^{-\beta (E_{k,i}+\mu_i^*)})\]\no\\
       &&=-\012 M^2_A A_0 ^2 -2 T  \int _k
        \[\ln (1+e^{-\beta (E_k-\mu^*)})\right.\no\\
       &&
       ~~~~~~~~ \left. +\ln (1+e^{-\beta (E_k+\mu^*)})\],
\eea where $``2"$ represents the hyperfine-spin degenerate factor
    of the fermions system.
Furthermore, the self-consistency condition or the tadpole diagram
    with the boson self-energy for the full fermion propagator leads to
\bea\label{electron-field}
        A_0=-\0{g}{M^2_A}n=-\0{2\pi a_{eff}}{g m}n,
\eea from which the effective (local) chemical potential
    $\mu ^*$ is defined with a gauge invariant manner
\bea\label{chemical}
    \mu^*&&\equiv\mu+\mu _{I}
    =\mu -\0{2\pi a_{eff}}{m} n,
\eea where $\mu$ is the global chemical potential. The fermions
particle number (electric charge number) density is \bea
    n=2\int _k \[f-\overline{f}\].
\eea

\subsection{Thermodynamics near the strongly interacting
unitary limit regime}

 With the external field counterterm instantaneous approximation
formalism of the interior correlating effects,
    one obtains the general analytical expressions for the energy density
    functional from the effective
    potential \eq{potential} with the
    thermodynamics relation
    $\epsilon =\0\Omega V  +\mu  n+T\0SV$
\bea\label{energy}
    \epsilon &&=\0{\pi a_{eff}}{m} n^2 +2\int_k E_k \[f +{\bar f}\],
\eea
    and pressure $P=-\Omega/V$
\bea\label{pressure}
    P=\0{\pi a_{eff}}{m} n^2+\0{2 }{d} \int_k \0{\vk^2}{E_k}\[f +{\bar
    f}\],
\eea with the entropy density \bea\label{entropy} \0SV=-2 \int_k
    \[f\ln f+(1-f)\ln \[1-f\]+(f\to \overline{f})\]\no\\
        =\01T \(2 \int _kE_k (f+{\bar f})+\0{2}d \int _k\0{{\bf k}^2}{E_k}(f+{\bar f}) -\mu^*n\).\no\\
\eea

In \eq{energy} and
    \eq{pressure}, the first terms are directly related to the interaction and/or quantum fluctuation contributions. In the weak coupling limit,
    the bare vacuum interaction strength/scattering
    length $a_{eff}\to a$ can recover the lowest order conventional mean field theory which neglects
    the quantum fluctuation contributions.
The second terms in \eq{energy} and \eq{pressure} as well
     as the entropy density \eq{entropy}
    appear as very much the analytical formalisms for the free Fermi-Dirac gas.
However, the correlating effects are also implicitly included
through
    the effective chemical potential esp. for $T\neq 0$.

The \eq{energy} and \eq{pressure} with $m^2_{{\tilde B}}=-\Pi
    _A^L(0,|\vp|\to 0)=\Pi
    _A^{00}(0,|\vp|\to 0)$ are the final results.
The remaining task is to give the numerical
        results for given temperature $T$ and density $n$ with the fermion mass $m$.
In this work,
        we limit ourselves to the $T=0$
        universal thermodynamics attempting to obtain the analytical results.

\subsection{Thermodynamics universality at $T=0$ for three-dimensions}

At $T=0$ and from the above general analytical expressions,
    one can obtain the energy density and pressure for finite scattering length $a$
\bea\label{energy-density}
    \epsilon &&=\0{\pi a }{m-\0{2 k_f E_fa}{\pi}}n^2+
    \0{(2k_f^2+m^2)k_fE_f-m^4\ln\0{k_f+E_f}{m}}{8\pi^2},
    \no\\
    P&&=\0{\pi a }{m-\0{2 k_f E_fa}{\pi}}n^2+\0{(2k_f^2-3m^2)k_fE_f+3m^4\ln\0{k_f+E_f}{m}}{24\pi^2}.\no\\
\eea In the non-relativistic limit with $T=0$,
    the analytical expression for
    the energy density can recover Steele's main result obtained
    within the effective theory framework\refr{Steele}.
In deed, the non-relativistic negative Debye/Thomas-Fermi mass
squared can readily approach the non-relativistic power counting
result of Steele with the generalized coupled Dyson-Schwinger
equations(through the full anti-screened vector boson
¡°propagator¡±) as indicated by Fig.\ref{fig1}. The non-trivial
physics occurs at the pole of the first term in
\eq{energy-density}, i.e., in the repulsive BEC regime with $m-2
k_f E_f a/\pi= 0 \to 2k_fa/\pi \approx 1$ as pointed out in Refs.
\refr{Pethick, Steele}.

Due to the universal properties at $T=0$ in the dilute unitary
    limit,
    the energy density can be scaled as\refr{cond-mat/0404687}
\bea
 \varp (n)&&=\035 \varp_f
    F(\01{k_f a})\no\\
    &&=\035 \varp_f\(\xi -\0{\zeta}{k_f a}-\053 \0{v}{k_f^2
    a^2}+0(\01{(k_fa)^3})\),
 \eea
where $\varp_f={k_f^2}/{(2 m) }$ is the Fermi kinetic energy.
 Applying the Taylor series expansion of $\epsilon$ \eq{energy-density}
according to
    $1/{(ak_f)}$ as well as $k_f$ by keeping only up to the lowest order of ${k_f}/{m}$,
    the corresponding universal coefficients
    $\xi=4/9\doteq 0.44, \zeta={5\pi}/{18}\doteq 0.87, v={\pi ^2}/{12}\doteq 0.82 $
    are exactly consistent with those $\xi\approx 0.44,\zeta\approx 1,v\approx 1$ of
    Refs.\refr{cond-mat/0404687,physics/0303094}
    obtained with Monte Carlo calculation.
Especially, one of them with much attention is $\xi=4/9$. The
recent quantum Monte Carlo calculation also gives the same
result\refr{cond-mat/0608154}.

In the unitary limit, the strong fluctuation and correlation
effects not only do manifest themselves in the bulk properties but
also modify the
    quasi-particle properties in a substantial way from the viewpoint of the universal
    thermodynamics.
At unitary $|a|=\infty$,
     the energy gap $\Delta$ is
    derived from the total energy density with the odd-even
    staggering (OES)
\bea\label{delta} \Delta= f_{sw} \0{2 k_f^2+3 m^2 -3 m E_f}{6
    E_f},
\eea with $f_{sw}$ being the Fermi-Dirac statistical weight
    factor. In the non-relativistic limit with the factor
    $f_{sw}=5/3$, the analytical result is
    $\Delta/\varepsilon _f ={5}/{18}$,
which is reasonably consistent with theoretical result $\0{8}{e^2}
    \varp _fe^{-\0{\pi}{2 k_f a}}\sim 0.40 \varp_f$ obtained through
    the BCS model but with the effective scattering length.
The corresponding phase transition critical temperature is still
approximated by the BCS relation
    $T_c={e^{\gamma}\Delta}/{\pi }$ with $\gamma$ being the
    Euler-Gamma constant
\bea T_c\approx 0.157 T_f,\eea as given in Ref.\refr{chen2006}.
This is in reasonable agreement with the updating
    results\refr{Bulgac2005,cond-mat/0608154,Burovski2006,Haussmann2006}.
%Of course, it is wondered whether the conventional quasi-particle
%concept of
%    the weak coupling degenerate fermions can be transplanted directly
%    to this strongly coupling limit problem, which deserves further study.

The pressure is $P=1/6P_{FG}$ in the unitary limit,
    from which one can find the sound speed is reduced remarkably
\bea
    v=\sqrt{\016}v_{FG}=\0{\sqrt{2}}6v_f,
\eea where $v_{FG}^2=1/3v_f^2$ is the sound speed squared for the
    ideal Fermi gas with the Fermi velocity $v_f={k_f}/{m}$. For the
    ideal fermion gas,
    the ratio of pressure to energy density is well known ${2}/{3}$.
For unitary fermion gas, this ratio is found to be changed to
    $1/4$ for the non-relativistic occasion due to the strongly correlating effects.
This is different from that obtained with the universal hypothesis
based on assuming the scaling property in terms of the zero-energy
bound state. This difference can be attributed to the implicit
pairing
    correlation contribution to the binding energy in the strongly
    coupling limit\refr{note1}.
The simultaneous experimental detection for the universal
    coefficient $\xi$ and the sound speed can judge this dilemma.
We
    also note that similar conclusion can be seen in the analytical
    Dyson-Schwinger attempt of Ref.\refr{Haussmann2006}.

\section{$D$-dimensions fermions universal properties at
unitarity}\label{sec4}

 Although the unitary limit issue is proposed for the
    three-dimensions physics representing for such as the inner neutron star crust physics
    in the low energy strongly interacting nuclear many-body theory context,
    it is very interesting to visit the $d$-dimensions occasion of the
    general Fermi-Dirac statistics as a toy model in the unitary limit
    by applying the Thomson Problem counterterm analytical method.

Concentrating on the realistic physics realized in the
    experimental environments,
    the non-relativistic analytical results reduced from the general expressions \eq{energy} and \eq{pressure} in
    the unitary limit with the Thomson Problem counterterm approach are given below.
The kinetic energy density contribution for $d$-dimensions
non-relativistic fermion gas is \bea \varepsilon
_{ideal}=\0{2d}{\(2\sqrt{\pi}\)^d(d+2)}
    \0{ \varp _fk_f^d}{\Gamma (\0d2 +1)},
\eea while the pressure is \bea P
    _{ideal}=\0{4}{\(2\sqrt{\pi}\)^d(d+2)}
    \0{ \varp _fk_f^d}{\Gamma (\0d2 +1)}.
\eea The negative correlation energy is derived to be \bea
    \varp_{int}=-\0{2}{\(2\sqrt{\pi}\)^dd}
    \0{ \varp _fk_f^d}{\Gamma (\0d2 +1)}. \eea

Analogously to the discussion for
     three-dimensions,
     one can obtain the general universal dimensionless
    coefficient $\xi$ expression according to dimension $d$
\bea
    \label{unifinal} \xi =f_{sw} \0{(d-2)(d+1)}{d(d+2)}=\0{(d-2)(d+1)}{d^2}.
\eea With the OES method for the $d$-dimensions unitary fermions
gas,
    one can have the extended $S$-wave energy gap
\bea
    \label{oes} \Delta/\varp_f =f_{sw}\0{d-2}{2d}=\0{(
    d-2)(d+2)}{2d^2}.
\eea In above expressions, $f_{sw}={(d+2)}/{d}$ is the
$d$-dimensions Fermi-Dirac statistical weight factor in the
non-relativistic limit.
 It is worthy noting that the ratio of pressure to
energy
    density is changed from $2/d$ of the ideal fermion gas to
    $1/{(1+d)}$ of that in the unitary limit.

From \eq{unifinal}, one can find that $\xi $ will be equal to zero
    at $d=2$.
This corresponds to the infinity $m^*=\infty$ of the
    effective fermion mass.
At this specific two-dimensions,
    the pressure as well as entropy density is also equal to zero,
    which is quite similar to the BEC phenomena for the ideal boson
    gas at three-dimensions.
In this specific occasion,
    the sound speed will approach to zero
    in terms of the Landau Fermi liquid theory.
Furthermore, the energy gap will be zero in the meantime with the
    OES result \eq{oes} of the two-dimensions.
This result indicates the low two-dimensions unitary limit
    thermodynamics properties are exotic.

\begin{figure}[ht]
        \centering
        \psfig{file=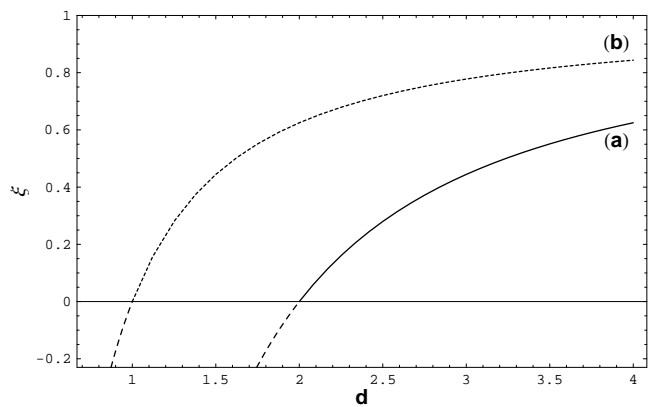,width=8.5cm,angle=-0}
        \caption{
        \small
The universal coefficient $\xi$ versus dimension $d$. The lower
solid line $(a)$ is for the non-relativistic \eq{unifinal}, while
the upper dotted line $(b)$ is for the ultra-relativistic limit
\eq{relativistic}. The dashed parts correspond to the instability
region.}\label{fig2}
\end{figure}

In the strongly coupling limit,
    the essential characteristic $m^*\to
    \infty$ is in line with the strongly coupled two-dimensions electrons discussed in the
    literature\refr{zhang2005}.
The zero energy gap is so similar to the vanishing of the extended
    $S$ pairing correlation etc. with the increase of \textit{repulsive} interaction strength in the two-dimensions Hubbard
    Model with quantum Monte Carlo
    study\refr{Zhang1997}.
It is worthy noting that the behavior of $\xi $ according to the
    spatial dimensions $d$ as indicated by Fig.\ref{fig2} is different from
    those found in the recent works\refr{Nussinov}.

In Fig.\ref{fig2}, we also give the ultra-relativistic limit
    result for completeness with the explicit analytical
    expressions
\bea\label{relativistic}
    \xi' &&=f'_{sw}\0{(d-1)(2d+1)}{2d(d+1)}=\0{(d-1)(2d+1)}{2 d^2},\eea
and \bea\label{relativistic-gap}
    \Delta'/\varp'_f &&=f'_{sw}\0{(d-1)}{2d}=\0{(d-1)(d+1)}{2
    d^2}. \eea
It is worthy noting that in this opposite limit the corresponding
correction of the Fermi kinetic energy is the magnitude of the
Fermi momentum $k_f$, i.e., $\varp'_f\rightarrow k_f$. The prime
symbol has been adopted in order to avoid the confusion with the
non-relativistic results. In
\eqs{relativistic}-(\ref{relativistic-gap}), the statistical
weight factor is $f'_{sw}={(d+1)}/d$. There exists a rummy duality
relation between the non-relativistic and the ultra-relativistic
results seen from above expressions.

From Fig.\ref{fig2}, one can clearly see that these solutions
    manifest the quantum Ising universal class characteristic of the strongly
    interacting fermions gas realized in the experimental
    environments\refr{Sachdev1999}.
For the non-relativistic occasion,
    the universal coefficient is $\xi<0$ for $d<2$ with the negative
    sound speed squared.
The negative sound speed squared means that there is no phase
    transition from the superfluidity state to normal state due to the
    spinodal instability of the superfluidity state. In the
    one-dimension, the strong fluctuation effects in the system will
    make the regions out of phase with each other and reduce the
    possibility of reaching the thermodynamics equilibrium.
This is consistent with the general phase transition theory, i.e.,
    there is not a phase transition for one-dimension.

With the analytical results such as \eq{unifinal} and \eq{oes} as
well as \eq{relativistic} and \eq{relativistic-gap}, one can also
see that the proportionality coefficients do depend only on the
spatial dimensions $d$ and are independent of the system details
in the unitary limit. This is of the genuine thermodynamics
universality characteristic associated with the critical phenomena
in terms of the general statistical physics.

\section{Summary}\label{sec5}

In conclusion,
    we have given the general analytical formulas of the energy density functional and pressure as well as entropy
    density at both finite temperature and density in terms of the
    relativistic quantum many-body formalism near the unitary limit regime.
The \textsl{interior} non-linear fluctuation/correlation effects
on
    the thermodynamics have been incorporated with an \textsl{external}
    field approximation scheme through a fictive opposite charged Thomson
    background.
This detour will allow us to go
    beyond the lowest order mean field theory or the naive perturbative expansions in an accurate way.

The $d$-dimensions universal thermodynamics in the unitary limit
    are discussed.
    It is found the thermodynamics properties of low dimension $d=2$
    are very interesting in the unitary limit, i.e., the universal
    dimensionless coefficient $\xi=0$ with simultaneous vanishing of
    energy density, pressure, energy gap and sound speed as well as entropy density.
In the non-relativistic limit, the effective fermion mass can
    approach to $m^*=\infty$,
    corresponding to the strongly coupled
    electrons in condensed matter physics.
This characteristic is quite similar to the well-known BEC of the
ideal boson
    gas at $T=0$ for three-dimensions. These $d$-dimensions solutions manifest the quantum Ising universal class characteristic of
the strongly interacting low temperature fermions gas physics.

In addition to the exact consistency of the obtained physical
    results with some existed theoretical analytical or simulation ones,
    this attempt with \textsl{the unknown side/Thomson Problem}
    as a potential quantum many-body nonperturbative arm to \textsl{solve the other
    unknown side} facilitates the concrete comparison of the
    non-relativistic  many-body methods and the Dirac phenomenology of
    the relativistic continuum formalism in terms of the fundamental in-medium Lorentz violation.

The essential features of the strongly coupled unitary gases may
be captured by the Lorentz violation with vector condensation
formalism. The key physics may be obtained by the external field
realization scheme of the interior correlations between the
strongly interacting particles through a fictive Thomson
background but still with the standard RHA and RPA techniques (or
the generalized coupled Dyson-Schwinger equations).

 \acknowledgments{The author
acknowledges the discussions with
    Profs. Jia-Rong Li and Lu Yu; colleagues' comments at the sQGP workshop at IOPP of CCNU, Aug 14-16,
    2006. He is also grateful to the beneficial communications with Drs. E. Burovski, F. Chevy, C. Chin,
    Jason Ho, C.J Pethick and S.-W Zhang. Supported by the scientific starting research
fund of CCNU and NSFC under grant No 10675052.}

\end{document}